\documentstyle[preprint,aps]{revtex} 

\newcommand{\be}{\begin{equation}}
\newcommand{\ee}{\end{equation}}
\newcommand{\bey}{\begin{eqnarray}}
\newcommand{\eey}{\end{eqnarray}}

\begin{document} 
\draft

\title{ Nonperturbative approach to a simple model with ultravioletly
divergent eigenenergies in perturbation theory }

\author{Wen-ge Wang}
\address{ 
Department of Physics, South-east University, Nanjing 
210096, China 
}
 
\date{\today}
\maketitle

\begin{abstract}
We study a simple quantized model for which perturbation
theory gives ultravioletly divergent results. We show
that when the eigen-solution problem of the Hamiltonian of
the model is treated nonperturbatively, it is possible for
eigenenergies of the Hamiltonian to be finite.

\end{abstract}

%\begin{multicols}{2}

\section{Introduction}
\label{Intro}

    Models in quantum field theories usually have the problem of
ultraviolet divergence in the framework of perturbation theory
(see, e.g., \cite{Itzy,Weinberg,MS}).
For renormalizable models, ultraviolet divergence can be
removed by renormalization techniques, but at the cost that
renormalized masses of particles can not be explained in the
framework of the theory.
On the other hand, non-renormalizable models are usually
rejected, since for them it is not clear at the present stage how to
obtain finite results for high order corrections in perturbation
theory.
However, rigorously to say, the break-down of perturbative
treatment to a model does not necessarily mean that the model cannot
give meaningful results when it is treated nonperturbatively.
The point here is that nonperturbative approach to
behaviors of states of ultravioletly divergent models
is generally quite difficult.
In particular, it is still not clear whether the Hamiltonian
of a model suffering ultraviolet divergence in the framework of
perturbation theory could have eigenstates with finite energies.
This problem is of interest, since, if the answer is positive,
then, the ultraviolet divergence may be avoidable in
nonperturbative treatment and finite energies of ground states
may be associated with masses of particles observed experimentally.

    In this paper we will show that the answer is indeed positive.
For this purpose, we study a simple quantized model, which gives
ultravioletly divergent results in the framework of perturbation theory.
The model has an interaction structure 
similar to that of QED in the number representation,
so that most of the arguments given
to it can be extended to the case of QED.
(In this paper, by interaction structure
in a representation, we mean the way in which 
basis states of the representation are coupled by
the interaction term of the Hamiltonian, that is, 
the structure of the non-zero off-diagonal elements of the Hamiltonian
matrix in the representation.)
For this reason, whether the
simple model has an appropriate form in configuration space
does not matter here.

    Similar to QED, the Hilbert space of the simple model
studied in this paper is infinite even
when the momentum space is cut off and discretized.
For the infinite Hilbert space, not all the theorems for
eigen-solutions of Hamiltonians in finite Hilbert spaces hold.
For example, eigenstates of the Hamiltonian of the simple model do not span
the whole infinite Hilbert space.
In particular, we find that eigenstates and eigenenergies of
the Hamiltonian can be expressed in terms of themselves,
and, as a result, the eigenenergies
can remain finite when the cut-off of momentum is taken off,
even if second order corrections to the eigenenergies in perturbation theory
is ultravioletly divergent.

    Concretely, this paper is organized in the following way.
In section \ref{sect2}, we introduce the quantized model,
discuss the structure of basis states 
and the structure of the Hamiltonian matrix in the basis states.
In section \ref{sect3}, we truncate the infinite Hilbert
space of the model,
so that obtain a series of finite Hilbert spaces, the limit of
which gives the infinite Hilbert space. For each truncated 
finite Hilbert space, we construct another set 
of basis states by making use of energy
eigenstates of another finite Hilbert space.
The Hamiltonian matrix in
the representation of the new set of basis states is quite
simple and can be diagonalized easily. 
Section \ref{sect4} is devoted
to discussions for eigenstates and eigenenergies  of the
Hamiltonian of the model in the infinite Hilbert space. We show that
it is possible for the eigenenergies to be finite, even if
perturbative treatment gives divergent results.
Conclusions and discussions are given in section \ref{sect5}.

\section{A simple quantized model}
\label{sect2}

    Since realistic models, such as the standard model, are complicated,
in this paper, as a first step to the method that is to be
developed for nonperturbative approach
to eigenstates and eigenenergies of Hamiltonians in quantum
field theories, we choose a model as simple as possible to study.
Such a model should have an interaction structure similar to
that of QED and may have ultraviolet divergence,
so that the method developed in this paper for the model
can be extended to treat realistic models, such as QED and
the standard model.
Since here we are interested in
properties of eigen-solutions of Hamiltonians only,
we are not to start from a classical Lagrangian expressed in configuration
space, but start from a Hamiltonian expressed in terms of
creation and annihilation operators for free fields.

    The simplest model satisfying the above requirements
is composed of a quantized fermion field and a quantized boson field
 in 1-dimensional momentum space. 
Denoting the creation and annihilation operators for a free fermion
 field with momentum $p$ and for a free boson field with momentum $k$
 as $b^{\dag }(p)$, $b(p)$, and $a^{\dag }
 (k)$, a(k), respectively, the Hamiltonian of the model is taken as
\be \label{H} H=H_f + H_b +H_I \ee
where
\bey  \label{Hf} H_f= \int p_0 b^{\dag }(p) b(p) dp \hspace{8cm}
\\ \label{Hb} H_b= \int k_0 a^{\dag }(k) a(k) dk     \hspace{8cm}
\\ \label{Hfbi} H_I= \int \left ( V(p_1, p_2,k) b^{\dag }(p_2) b(p_1)
a^{\dag }(k) + h.c. \right )
  \delta (p_1-p_2-k) dp_1 dp_2 dk
\eey
 with $p_0=|p| $ and $k_0=|k|$.
The operators $b^{\dag }(p)$ and $b(p')$ satisfy the usual
 anticommutation relations and 
 $a^{\dag }(k)$ and $a(k')$ satisfy the usual
commutation relations.   Here $\hbar $ and $c$ are taken to be
unit, $\hbar = c =1$.
The interaction structure of this Hamiltonian in the number representation,
given by the expression of $H_I$ in Eq.~(\ref{Hfbi}), 
is clearly similar to (although simpler than) that of QED.

    For the sake of convenience in discussing properties of
the Hamiltonian $H$,
we discretize the momentum space and take a cut-off $\Lambda $,
that is, we take
\be \label{pi}  p =p_i = i \cdot \Delta p - \Lambda , \hspace{2cm}
 k =k_j = j \cdot \Delta p - \Lambda , \ee
where $i,j = 0,1, \ldots , N$ with $N=2\Lambda / \Delta p $.
The Hamiltonian of the model  
expressed in terms of summations over $p=p_i$ and $k=k_j$ is 
\be \label{H2} H(\Lambda )=H_f(\Lambda ) + H_b(\Lambda ) +H_I(\Lambda ), \ee
where 
\bey \label{Hf2} H_f(\Lambda )= \sum_p p_0 b^{\dag }(p) b(p) \hspace{5.3cm}
\\ \label{Hb2} H_b(\Lambda )= \sum_k k_0 a^{\dag }(k) a(k) \hspace{5.3cm}
\\ \label{Hfbi2} H_I(\Lambda )= \sum_{p_1,p_2} \left ( V(p_1, p_2)
b^{\dag }(p_2) b(p_1) a^{\dag }(p_1-p_2) + h.c. \right ).
\eey
In the limit of $\Delta p \to 0$ and $\Lambda \to \infty $,
$H(\Lambda )$ becomes the $H$ in Eq.~(\ref{H}). 
In this paper we assume that with finite cut-off $\Lambda $
the model is free from divergence. 

    Since $H_I(\Lambda ) a^{\dag }(k)|0\rangle =0$, where $|0\rangle $ is
the vacuum state, states $a^{\dag }(k)
|0\rangle $, denoted by $|s_k\rangle $ after normalization, are eigenstates
of the Hamiltonian $H(\Lambda )$. We are not interested in
this kind of trivial eigenstates here. What we are interested in
are fermion-type eigenstates. The simplest
fermion-type eigenstates of the Hamiltonian $H(\Lambda )$ are 
in the Hilbert space spanned by the state $b^{\dag }(p)|0\rangle $,
denoted by $|f_p\rangle $ after normalization, and all the
states that can be coupled to 
$|f_p\rangle $ by $H^m_I(\Lambda )$, the product of $m$
$H_I(\Lambda )$,  with $m=0,1,2,\ldots $.
It is this Hilbert space that we are to study in this paper,
which will be denoted by $L_{\infty }(p,\Lambda )$.

    Concretely to say, basis states of the Hilbert space
$L_{\infty }(p,\Lambda )$ can be taken as
\be \label{bas0} |f_{p-k_1- \ldots -k_m}s_{k_1} \ldots s_{k_m}\rangle
\equiv N_{p,k_1,\ldots ,k_m } b^{\dag }(p-k_1 - \ldots -k_m)
a^{\dag }(k_1) \ldots a^{\dag }(k_m) |0\rangle \ee
for $m=0,1,2,\ldots $ (the case for $m=0$ is just $|f_p\rangle $),
where $N_{p,k_1,\ldots ,k_m}$ are normalization coefficients.
Noticing that the basis states
$|f_{p-k_1- \ldots -k_m}s_{k_1} \ldots s_{k_m}\rangle $ for all possible
$k_1, \ldots ,k_m$  can be coupled
to the state $|f_p\rangle $ by $H_I^m(\Lambda )$,
we will denote the set of them
by $\{ H_I^m(\Lambda )|f_p\rangle \}$ in what follows.
Then, the basis states of the infinite Hilbert
space $L_{\infty }(p,\Lambda )$ in Eq.~(\ref{bas0})
are elements of the following sets
\be \{H_I^0(\Lambda )|f_p\rangle =|f_p\rangle \},
\{ H_I(\Lambda )|f_p\rangle \}, \{H_I^2(\Lambda )|f_p\rangle \},
\{ H_I^3(\Lambda )|f_p\rangle \}, \ldots . \label{basis0} \ee
In some cases in the following sections, for brevity, 
instead of the expression in Eq.~(\ref{bas0}),
we use $|\xi _{i_m}(m,p,\Lambda )\rangle $
to denote basis states in the set $\{ H_I^m(\Lambda )|f_p\rangle \}$,
i.e., we use $i_m$ to denote $(k_1,\ldots , k_m)$.
For example, $|\xi _{i_0}(0,p,\Lambda )\rangle $ indicates
$|f_p\rangle $ with $i_0 =1$, $|\xi _{i_1}(1,p,\Lambda )\rangle $
indicates $|f_{p-k_1}s_{k_1}\rangle $ with $i_1 = k_1$, and so on. 

    The interaction structure of the Hamiltonian $H(\Lambda )$
in the basis states in Eq.~(\ref{bas0}), i.e., 
the structure of the non-zero off-diagonal elements of 
the Hamiltonian matrix in the basis states,  has an
interesting tree structure: 
The basis state $|f_p\rangle $ is coupled to basis states in the set
$\{ H_I(\Lambda )|f_p\rangle \}$ only; basis states in the set
$\{ H_I(\Lambda )|f_p\rangle \}$ are coupled to basis states in the sets
$\{ |f_p\rangle \}$ and $\{ H_I^2(\Lambda )|f_p\rangle \}$ only; $\ldots $;
basis states in the set $\{ H_I^m(\Lambda )|f_p\rangle \}$
are coupled to basis states in the sets
$\{ H_I^{m-1}(\Lambda )|f_p\rangle \}$
and $\{ H_I^{m+1}(\Lambda )|f_p\rangle \}$ only; $\ldots $. 
It is easy to show that, with this structure of the Hamiltonian
matrix, when the coupling strength $V(p_1,p_2)$ in
Eq.~(\ref{Hfbi2}) is strong enough, second order corrections
to the eigenenergies of the Hamiltonian $H(\Lambda )$ are ultravioletly
divergent in perturbation theory when the cut-off $\Lambda $
approaches to infinity.

\section{ Truncated Hilbert space and $\psi_s$ representation}
\label{sect3}

    The Hilbert space $L_{\infty }(p,\Lambda )$
spanned by the basis states in the sets
given in (\ref{basis0}) is infinite, although the momentum
space has been cut off by $\Lambda $.
Since the problem of eigen-solutions of
a Hamiltonian in an infinite Hilbert space
is more difficult than that in a finite Hilbert space
and many theorems in the latter case are invalid in the former case,
in this section we truncate the Hilbert space to
a series of finite ones and
discuss properties of the eigenstates of the Hamiltonian
$H(\Lambda )$ in the
truncated finite Hilbert spaces. Then, in the next section, we
discuss what we could have when the truncated finite Hilbert spaces
resume the infinite Hilbert space $L_{\infty }(p,\Lambda )$.

\subsection{Truncated finite Hilbert spaces }
\label{sect3.1}

    A truncated finite Hilbert space, denoted by $L_{n}(p,\Lambda )$,
is spanned by basis states in a
set $A(n,p,\Lambda )$ defined by
\be A(n,p,\Lambda ) = \{|f_p\rangle \} \bigcup
\{ H_I(\Lambda )|f_p\rangle \} \bigcup \ldots \bigcup 
\{ H_I^n(\Lambda )|f_p\rangle \}. \label{Apn} \ee
When $n$ goes to infinity, the truncated Hilbert space
$L_{n}(p,\Lambda )$ will become the infinite
Hilbert space $L_{\infty }(p,\Lambda )$.

    Normalized eigenstates and the corresponding eigenenergies of
the Hamiltonian $H(\Lambda )$ in the truncated finite Hilbert
space $L_{n}(p,\Lambda )$ are denoted
by $|\psi_{\alpha _n}(n,p,\Lambda )\rangle $
and $E_{\alpha _n}(n,p,\Lambda )$, respectively, 
\be H(\Lambda )|\psi_{\alpha _n}(n,p,\Lambda )\rangle = E_{\alpha _n}
(n,p,\Lambda ) |\psi_{\alpha _n}(n,p,\Lambda )\rangle .\label{Ealpha} \ee
These states $|\psi_{\alpha _n}(n,p,\Lambda )\rangle $
also span the Hilbert space $L_n(p,\Lambda )$ and
can be expanded in the basis states in the set $A(n,p,\Lambda )$,
\be \label{psi-xi} |\psi_{\alpha _n}(n,p,\Lambda )\rangle  =
\sum_{m=0}^n \sum_{i_m} C_{\alpha _n, i_m}(n,m,p,\Lambda )
|\xi_{i_m}(m,p,\Lambda )\rangle , \ee
where $C_{\alpha _n, i_m}(n,m,p,\Lambda )$ are expanding coefficients.

    The set $A(n,p,\Lambda )$ has an interesting structure:
It can be expressed by making use of the
sets $A(n-1,p-k,\Lambda )$. To see this,
first note that the product of two basis states $|s_k\rangle $
and $|f_{p-k-k_1-\ldots -k_{m}}s_{k_1} \ldots s_{k_m}\rangle $ is
\be |s_k\rangle \cdot 
|f_{p-k-k_1-\ldots -k_{m}}s_{k_1} \ldots s_{k_{m}}\rangle =
|f_{p-k-k_1-\ldots -k_{m}}s_{k_1} \ldots s_{k_{m}} s_k \rangle .
\label{pb} \ee
From Eq.~(\ref{bas0}), it is easy to see that 
each basis state in the set $\{ H_I^m(\Lambda )|f_p\rangle
\}$ with $m \ge 1$ can be expressed as the product of
a basis state $|s_k\rangle $ and a basis state in the set 
$\{ H_I^{m-1}(\Lambda )|f_{p-k} \rangle \} $.
Then, denoting 
the set of the product of the basis states
$|s_k\rangle $ and the basis states in the set $A(n-1,p-k,\Lambda )$
for all possible $k$ as $B(n-1,p,k,\Lambda )$,
\be \label{B} B(n-1,p,k,\Lambda ) =
\{ |s_k\rangle \cdot |\xi \rangle : \ \ {\rm for} \  |\xi \rangle \in
A(n-1,p-k,\Lambda ) \ \ {\rm and \ all \ possible} \ k \}, \ee
the set $A(n,p,\Lambda )$ can be reexpressed as 
\be A(n,p,\Lambda ) = \{ |f_p\rangle \} \bigcup_k
B(n-1,p,k,\Lambda ). \label{AB} \ee
In the next subsection, we show that this structure of the
set $A(n,p,\Lambda )$ enables us to construct another set of
basis states, in which the Hamiltonian matrix can be diagonalized
explicitly.

\subsection{$\psi _s$-representation in truncated finite Hilbert spaces }
\label{sect3.2}

    Making use of the definition of product of
basis states given
in Eq.~(\ref{pb}) and the expansion of eigenstates 
in Eq.~(\ref{psi-xi}), it is easy to get
$|s_k \rangle \cdot | \psi_{\alpha _{n-1}}(n-1,p-k,\Lambda )\rangle $,
the product of a state
$|s_k\rangle $ and an eigenstate
$|\psi_{\alpha _{n-1}}(n-1,p-k,\Lambda )\rangle $
in the truncated Hilbert space $L_{n-1}(p-k,\Lambda )$, 
which will be denoted by
$|s_k \psi_{\alpha _{n-1}}(n-1,p-k,\Lambda )\rangle $,
\be \label{skp} |s_k \psi_{\alpha _{n-1}}(n-1,p-k,\Lambda )\rangle  =
\sum_{m=0}^{n-1} \sum_{i_m} C_{\alpha _{n-1}, i_m}(n-1,m,p-k,\Lambda )
|s_k\rangle \cdot |\xi_{i_m}(m,p-k,\Lambda )\rangle . \ee
These  states 
$|s_k \psi_{\alpha _{n-1}}(n-1,p-k,\Lambda )\rangle $
span the same Hilbert space as the states in the set
$B(n-1,p,k,\Lambda )$ in Eq.~(\ref{B}).
Therefore, these states together with the state $|f_p\rangle $
also span the truncated Hilbert space $L_n(p,\Lambda )$.

    Now we show that in the limit of $\Delta p \to 0$,  the states 
$|s_k \psi_{\alpha _{n-1}}(n-1,p-k,\Lambda )\rangle $ are
orthogonal to each other and together with $|f_p\rangle $ form another
set of orthogonal basis states for the truncated Hilbert space
$L_n(p,\Lambda )$. 
To prove this, we rewrite 
$|\psi_{\alpha _{n-1}}(n-1,p-k,\Lambda )\rangle $
in the form
\be |\psi_{\alpha _{n-1}}(n-1,p-k,\Lambda )\rangle
=\sum_{m=0}^{n-1} |h_{\alpha _{n-1}}(n-1,m,p-k,\Lambda )\rangle ,
\label{ha} \ee
where
\be \label{h-xi} |h_{\alpha _{n-1}}(n-1,m,p-k,\Lambda )\rangle  =
\sum_{i_m} C_{\alpha _{n-1}, i_m}(n-1,m,p-k,\Lambda )
|\xi_{i_m}(m,p-k,\Lambda )\rangle , \ee
for example,
\bey \label{h-xi0} |h_{\alpha _{n-1}}(n-1,0,p-k,\Lambda )\rangle  =
C_{\alpha _{n-1},1}(n-1,0,p-k,\Lambda )
|f_{p-k}\rangle  \hspace{1.65cm}
\\ \label{h-xi1} |h_{\alpha _{n-1}}(n-1,1,p-k,\Lambda )\rangle  =
\sum_{k_1} C_{\alpha _{n-1},k_1}(n-1,1,p-k,\Lambda )
|f_{p-k-k_1} s_{k_1}\rangle .  \eey
The states $|h_{\alpha _{n-1}}(n-1,m,p-k,\Lambda )\rangle  $
are projections of the state 
$|\psi_{\alpha _{n-1}}(n-1,p-k,\Lambda )\rangle $ in the subspaces
spanned by states in the sets $\{H_I^m(\Lambda )|f_{p-k}\rangle \}$,
respectively,
therefore, $|h_{\alpha _{n-1}}(n-1,m,p-k,\Lambda )\rangle  $
with different $m$ are orthogonal to each other.

    Making use of Eq.~(\ref{ha}), we have
\be \langle s_{k'} \psi_{\alpha _{n-1}'}(n-1,p-k',\Lambda )
 |s_k \psi_{\alpha _{n-1}}(n-1,p-k,\Lambda )\rangle
= \sum_{m=0}^{n-1} I_m,    \label{psiI} \ee
where
\be I_m=  \langle s_{k'} h_{\alpha _{n-1}'}(n-1,m,p-k',\Lambda )
 |s_k \psi_{\alpha _{n-1}}(n-1,p-k,\Lambda )\rangle
\label{Im} \ee
with $|s_{k'} h_{\alpha _{n-1}'}(n-1,m,p-k',\Lambda )\rangle 
=|s_{k'}\rangle \cdot | h_{\alpha _{n-1}'}(n-1,m,p-k',\Lambda )\rangle $
defined in the same way as 
$ |s_k \psi_{\alpha _{n-1}}(n-1,p-k,\Lambda )\rangle$
in Eq.~(\ref{skp}).
It is easy to verify that
\bey \label{I0} I_0= C^*_{\alpha _{n-1}',1}(n-1,0,p-k',\Lambda )
C_{\alpha_{n-1}, 1}(n-1,0,p-k,\Lambda ) \delta_{k k'}
\hspace{1.5cm}
\\ \nonumber I_1= 
C^*_{\alpha_{n-1}', k}(n-1,1,p-k',\Lambda )
C_{\alpha_{n-1}, k'}(n-1,1,p-k,\Lambda ) 
\hspace{2cm}
\\ \label{I1} +\sum_{k_1} C^*_{\alpha_{n-1}', k_1}(n-1,1,p-k',\Lambda )
C_{\alpha_{n-1}, k_1}(n-1,1,p-k,\Lambda ) \delta_{k k'}.
\eey
Here we assume that in the limit of $\Delta p \to 0$, there is no
singularity in the coefficients 
$C_{\alpha_{n-1}, i_m}(n-1,m,p-k,\Lambda )$ for fixed $n-1$ and $m$,
i.e., $C_{\alpha_{n-1}, i_m}(n-1,m,p-k,\Lambda )$ is a smooth
function of $i_m$. 
Then, since
\be \sum_{k_1}
|C_{\alpha_{n-1}, k_1}(n-1,1,p-k,\Lambda )|^2 < 1, \label{norC} \ee
when the interval $\Delta p$ for discretizing the momentum $k_1$
goes to zero, the value of
$C^*_{\alpha_{n-1}', k}(n-1,1,p-k',\Lambda )
C_{\alpha_{n-1}, k'}(n-1,1,p-k,\Lambda ) $ should approach to zero,
and  Eq.~(\ref{I1}) gives 
\be \lim_{\Delta p \to 0}
I_1 =\sum_{k_1} C^*_{\alpha_{n-1}', k_1}(n-1,1,p-k',\Lambda )
C_{\alpha_{n-1}, k_1}(n-1,1,p-k,\Lambda ) \delta_{k k'}.
\label{I12} \ee
Similar results can also be obtained for the other $I_m$, and
finally we have
\bey \nonumber        \lim_{\Delta p \to 0}
\langle s_{k'} \psi_{\alpha _{n-1}'}(n-1,p-k',\Lambda )
 |s_k \psi_{\alpha _{n-1}}(n-1,p-k,\Lambda )\rangle  \hspace{5cm}
\\ =\sum_{m=0}^{n-1} \sum_{i_m}
C^*_{\alpha_{n-1}', i_m}(n-1,m,p-k',\Lambda ) 
C_{\alpha_{n-1}, i_m}(n-1,m,p-k,\Lambda ) \delta_{k k'}
=\delta_{\alpha_{n-1} \alpha_{n-1}'} \delta_{kk'}.
\label{psior} \eey
That is to say, 
the states $ |s_k \psi_{\alpha _{n-1}}(n-1,p-k,\Lambda )\rangle $
are normalized and orthogonal to each other.
Therefore, these states and the state $|f_p\rangle $ form another
set of normalized orthogonal basis states for the truncated
Hilbert space $L_n(p,\Lambda )$. The representation
given by this set of basis states in the limit of
$\Delta p \to 0$, will be termed
$\psi_s$-{\it representation} in what follows.

    The Hamiltonian $H(\Lambda )$ has a quite simple matrix form in
the $\psi_s$-representation of the Hilbert space $L_n(p,\Lambda )$.
In fact, similar to Eq.~(\ref{psior}),
one can verify that
\bey \nonumber \lim_{\Delta p \to 0} 
\langle s_{k'} \psi_{\alpha _{n-1}'}(n-1,p-k',\Lambda )
|H(\Lambda ) |s_k \psi_{\alpha _{n-1}}(n-1,p-k,\Lambda )\rangle \hspace{3cm}
\\ =\langle s_{k} \psi_{\alpha _{n-1}}(n-1,p-k,\Lambda )
|H(\Lambda ) |s_k \psi_{\alpha _{n-1}}(n-1,p-k,\Lambda )\rangle
\delta_{\alpha_{n-1} \alpha_{n-1}'} \delta_{kk'}.
\label{Hkk} \eey
Therefore, the non-zero off-diagonal elements of the
Hamiltonian matrix in the $\psi_s$-representation
are those connecting the state $|f_p\rangle $
and the states $|s_k \psi_{\alpha _{n-1}}(n-1,p-k,\Lambda )\rangle $ only.
Diagonalization of the Hamiltonian matrix with such a simple structure
is quite easy, which gives
\bey \nonumber 
E_{\alpha_n}(n,p,\Lambda )-\langle f_p|H(\Lambda )|f_p \rangle
\hspace{10cm}
\\  \label{energya} = \sum_{k,\alpha_{n-1} }
\frac{ \left | \langle f_p|H(\Lambda )|s_k
\psi_{\alpha _{n-1} }(n-1,p-k,\Lambda )\rangle \right | ^2}
{E_{\alpha_n}(n,p,\Lambda )
-\langle s_k \psi_{\alpha_{n-1} }(n-1,p-k,\Lambda )|H(\Lambda )|
s_k \psi_{\alpha_{n-1} }(n-1,p-k,\Lambda )\rangle } \eey
and 
\be  \label{psia}
|\psi_{\alpha_{n} }(n,p,\Lambda )\rangle =
D_{\alpha _n}(n,p,\Lambda )|f_p\rangle +
\sum_{k, \alpha_{n-1}} D_{\alpha _n , k \alpha_{n-1}}(n,p,\Lambda )
|s_k \psi_{\alpha_{n-1} }(n-1,p-k,\Lambda )\rangle , \ee
where
\be \label{D1} D_{\alpha _n ,k \alpha_{n-1}}(n,p,\Lambda )=
\frac{ \langle s_k \psi_{\alpha_{n-1}}(n-1,p-k,\Lambda )|H(\Lambda )
|f_p\rangle \cdot D_{\alpha _n}(n,p,\Lambda )}
{E_{\alpha_n}(n,p,\Lambda )
-\langle s_k \psi_{\alpha_{n-1} }(n-1,p-k,\Lambda )|H(\Lambda )|
s_k \psi_{\alpha_{n-1} }(n-1,p-k,\Lambda )\rangle }  \ee
and 
\bey \nonumber D_{\alpha _n}(n,p,\Lambda ) \hspace{14.4cm}
\\ \label{D0}= \left [ 1+ \sum_{k, \alpha_{n-1} }
\frac{ |\langle s_k \psi_{\alpha_{n-1}}(n-1,p-k,\Lambda )|H
(\Lambda )|f_p\rangle |^2}
{\left ( E_{\alpha_n}(n,p,\Lambda )
-\langle s_k \psi_{\alpha_{n-1} }(n-1,p-k,\Lambda )|H(\Lambda )|
s_k \psi_{\alpha_{n-1} }(n-1,p-k,\Lambda )\rangle \right ) ^2 }
\right ] ^{-\frac 12}. \eey

\section{Eigenenergies and eigenstates in infinite Hilbert space}
\label{sect4}

    When $n$ goes to infinity, the truncated Hilbert space $L_n(p,\Lambda )$ 
discussed in the previous section
will become the infinite Hilbert space $L_{\infty }(p,\Lambda )$.
However, eigenstates of the Hamiltonian $H(\Lambda )$
in the infinite Hilbert space cannot be
obtained by simply letting the index $n$ go to infinity in the eigenstates
$|\psi_{\alpha_{n} }(n,p,\Lambda )\rangle $ of the truncated
Hilbert space $L_n(p,\Lambda )$.
In fact, when $n$ becomes $n+1$, the number of eigenstates
in the truncated Hilbert space will become $N$ times larger
($N$ is the number of discretized momenta);
and despite of how large $n$ is, there exist eigenstates
$|\psi_{\alpha_{n} }(n,p,\Lambda )\rangle $ for which
the values of $|\langle h_{\alpha_n}(n,n,p,\Lambda )
|\psi _{\alpha_n}(n,p,\Lambda )\rangle |^2$ are not small,
i.e., there exist eigenstates whose projections in the
subspace spanned by states in the set $\{ H_I^n(\Lambda ) 
|f_p\rangle \} $ are not small.
Therefore, the definition for eigenstates of the
Hamiltonian $H(\Lambda )$ in the infinite Hilbert space $L_{\infty }
(p,\Lambda )$ should be treated more carefully.
In subsection \ref{sect4.1}, we give the definition
and discuss some properties of the eigenstates defined by it. 
In subsection \ref{sect4.2}, we show that eigenstates of the
Hamiltonian $H(\Lambda )$ in the infinite
Hilbert space $L_{\infty }(p,\Lambda )$ can be expressed
in terms of themselves, and the corresponding eigenenergies can remain
finite when the cut-off $\Lambda $ is taken off,
even if they are ultravioletly divergent in perturbation
theory. Subsection \ref{sect4.3} is devoted to a brief discussion
for evolution of states with time in the infinite Hilbert space.

\subsection{Energy eigenstates in infinite Hilbert space}
\label{sect4.1}

    A normalized eigenstate of the Hamiltonian $H(\Lambda )$ in the
infinite Hilbert space
$L_{\infty }(p,\Lambda )$, denoted by $|\phi_{\beta }(p,\Lambda )
\rangle $, with eigenenergy $E_{\beta }(p,\Lambda )$ is
defined in the following way: For each positive number $\epsilon $,
there exists a number $N(\epsilon )$ such that for each $n$ not smaller
than $N(\epsilon )$, there exists an eigenstate 
$|\psi_{\beta }(n,p,\Lambda )\rangle $ of the Hamiltonian $H(\Lambda )$
with eigenenergy $E_{\beta }(n,p,\Lambda )$
in the truncated Hilbert space $L_n(p,\Lambda )$ satisfying
\be 1-|\langle \psi_{\beta  }(n,p,\Lambda )|\phi_{\beta }
(p,\Lambda )\rangle |^2 < \epsilon \label{phid} \ee
and
\be |E_{\beta }(p,\Lambda ) - E_{\beta }(n,p,\Lambda )|
< \epsilon . \label{phied} \ee
Then, each eigenstate $|\phi_{\beta }(p,\Lambda ) \rangle $
is the limit of a series of states in the truncated Hilbert spaces
$L_n(p,\Lambda )$,
\be |\phi_{\beta }(p,\Lambda ) \rangle = \lim_{n \to \infty }
|\psi_{\beta }(n,p,\Lambda ) \rangle . \label{philim} \ee

    A property of an eigenstate
$|\phi_{\beta }(p,\Lambda ) \rangle $
is that its projection is less than $\epsilon $
in the subspace of the infinite Hilbert space spanned by states
in the sets $\{ H_I^{N(\epsilon )+1} \}$, $\{ H_I^{N(\epsilon )+2} \}$,
$\ldots $. In fact, substituting Eq.~(\ref{phid}) into the normalization
condition
\be \sum_{\alpha_{N(\epsilon )}} 
|\langle \psi_{\alpha_{N(\epsilon )}}
(N(\epsilon ),p,\Lambda )|\phi_{\beta }
(p,\Lambda )\rangle |^2
+ \sum_{m=N(\epsilon )+1}^{\infty } \sum_{i_m}  
|\langle \xi_{i_m}(m,p,\Lambda )|\phi_{\beta }
(p,\Lambda )\rangle |^2 =1, \label{norphi} \ee
we have 
\be \sum_{m=N(\epsilon )+1}^{\infty } \sum_{i_m}  
|\langle \xi_{i_m}(m,p,\Lambda )|\phi_{\beta }
(p,\Lambda )\rangle |^2 < \epsilon . \label{norphi2} \ee
Therefore, the projection of each of the eigenstates
$|\phi_{\beta }(p,\Lambda ) \rangle $ in the subspace spanned by
states in the set $\{ H_I^n(\Lambda )|f_p\rangle \} $ approaches
to zero when $n \to \infty $.
Due to this property of the eigenstates
$|\phi_{\beta }(p,\Lambda ) \rangle $, one can see that
for large $n$ and small $\epsilon $,
the number of the states 
$|\psi_{\beta }(n,p,\Lambda )\rangle $,
which are close to the eigenstates
$|\phi_{\beta }(p,\Lambda ) \rangle $,
must be much smaller than the total number of the states
$|\psi_{\alpha_{n}}(n,p,\Lambda )\rangle $ in the
truncated Hilbert space $L_n(p,\Lambda )$.
As a result, the eigenstates
$|\phi_{\beta }(p,\Lambda ) \rangle $ span a small
part of the infinite Hilbert space $L_{\infty }(p,\Lambda )$ only.

\subsection{Expressions of eigenenergies and eigenstates
in $\phi_s$-subspace}
\label{sect4.2}

    As discussed above, when $n \to \infty $,
not all the states $|s_k \psi_{\alpha _{n-1} }(n-1,p-k,\Lambda ) \rangle $,
but a fraction of them, namely,
$|s_k \psi_{\beta }(n-1,p-k,\Lambda ) \rangle $ have definite limit
\be  \lim_{n \to \infty }
|s_k \psi_{\beta }(n-1,p-k,\Lambda ) \rangle
= |s_k \phi_{\beta }(p-k,\Lambda ) \rangle . \label{ppp} \ee
The subspace of the Hilbert space
$L_{\infty }(p,\Lambda )$ spanned
by the states
\be \label{basis1} |f_p\rangle \ \ {\rm and } \ \
|s_k \phi_{\beta }(p-k,\Lambda )
\rangle \ee
for all possible $k$ and $\beta $
will be termed  $\phi_s$-{\it subspace}.
Since all the states
$|s_k \psi_{\alpha_{n-1} }(n-1,p-k,\Lambda )\rangle $
with $\alpha _{n-1} \ne \beta $ do not have definite limit when
$n$ goes to $\infty $,
the eigenstates $|\phi_{\beta }(p,\Lambda ) \rangle $
must lie in the $\phi_s$-subspace of the infinite Hilbert space
$L_{\infty }(p,\Lambda )$.
Therefore, if one diagonalizes the Hamiltonian $H(\Lambda )$
in the $\phi_s$-subspace, the eigenstates obtained
in the $\phi_s$-subspace must contain all the eigenstates 
$|\phi_{\beta }(p,\Lambda ) \rangle $ of the whole Hilbert space
$L_{\infty }(p,\Lambda )$.
As a result, in order to obtain the eigenstates 
$|\phi_{\beta }(p,\Lambda ) \rangle $, one does not
need to diagonalize the Hamiltonian $H(\Lambda )$ in the whole
Hilbert space $L_{\infty }(p,\Lambda )$, but diagonalization
in the $\phi_s$-subspace is enough. 

    Similar to Eq.~(\ref{psior}), 
one can show that the states in (\ref{basis1}) 
are normalized and orthogonal to each other.
Furthermore, the Hamiltonian matrix in 
the $\phi_s$-subspace has a structure similar to
that in the $\psi_s$-representation of the Hilbert space
$L_n(p,\Lambda )$, i.e., there is coupling between the state
$|f_p\rangle $ and each of the states
$|s_k \phi_{\beta }(p-k,\Lambda )\rangle $, but the coupling
between each two of the states $|s_k \phi_{\beta }(p-k,\Lambda )\rangle $
approaches to zero when $\Delta p \to 0$.
Then, similar to Eqs.~(\ref{energya}) and (\ref{psia}), 
in the limit of $\Delta p \to 0$, 
one can obtain expressions of the eigenstates and eigenenergies
of the Hamiltonian $H(\Lambda )$ in the $\phi_s$-subspace.
In particular, for
eigenenergies and eigenstates of the Hamiltonian
$H(\Lambda )$ in the whole
Hilbert space $L_{\infty }(p,\Lambda )$, after simplification we have 
\bey  \label{energyb}
E_{\beta }(p,\Lambda )
=p_0+ {\sum_{k,\beta ' }} '
\frac{ \left | \langle f_p|H(\Lambda )|f_{p-k} s_k 
\rangle \right | ^2 \cdot |d_{\beta '}(p-k,\Lambda ) |^2 }
{E_{\beta }(p,\Lambda ) -k_0 -E_{\beta '}(p-k,\Lambda ) }
\hspace{0.5cm}
\\  \label{psib}
|\phi_{\beta }(p,\Lambda )\rangle = D_{\beta }(p,\Lambda )|f_p\rangle
+ {\sum_{k, \beta '}}' D_{\beta , k \beta '}(p,\Lambda )
|s_k \phi_{\beta '}(p-k,\Lambda )\rangle , \eey
where
\bey \label{D1b} D_{\beta ,k \beta '}(p,\Lambda )=
\frac{ \langle f_{p-k} s_k |H(\Lambda )|f_p\rangle
d_{\beta '}(p-k,\Lambda )}{E_{\beta }(p,\Lambda )
- k_0 - E_{\beta '}(p-k,\Lambda )} D_{\beta }(p,\Lambda )
\hspace{1.5cm}
\\ \label{D0b} D_{\beta }(p,\Lambda )= \left [ 1+ {\sum_{k, \beta ' }}'
\frac{ \left | \langle f_p|H(\Lambda )|f_{p-k} s_k 
\rangle \right | ^2 \cdot |d_{\beta '}(p-k,\Lambda ) |^2 }
{ \left ( E_{\beta }(p,\Lambda )
-k_0 -E_{\beta '}(p-k,\Lambda ) \right ) ^2 }
\right ] ^{-\frac 12} \eey
\be  d_{\beta '}(p-k,\Lambda ) =
\langle \phi_{\beta '}(p-k,\Lambda )|f_{p-k} \rangle 
\label{Cbeta} \hspace{4cm} \ee
and the primes over the summations mean that when $k_0=0$
the index $\beta '$ is not equal to $\beta $.
Note that the right hand sides of Eqs.~(\ref{energyb}) and
(\ref{psib}) contain the eigenenergies and eigenstates themselves.

    Now let us compare the expression of the eigenenergy $E_{\beta }
(p,\Lambda )$ on the right hand side of Eq.~(\ref{energyb}) and the second
order correction for it in perturbation theory, which is 
\be E_{\beta }^{(2)}(p,\Lambda ) = {\sum_{k }}' \frac{ \displaystyle
|\langle f_p |H_I(\Lambda ) |f_{p-k} s_k \rangle |^2 }{\displaystyle
p_0 -|p-k| - k_0 }. \label{E2} \ee
The main difference between the right hand side of
Eq.~(\ref{energyb}) and the right hand side of Eq.~(\ref{E2})
is that there is a term
$|d_{\beta '}(p-k,\Lambda ) |^2$ in the numerator of
the right hand side of Eq.~(\ref{energyb}).
Equation (\ref{Cbeta}) shows that this term is less than one.
Let us consider the case that the summation on the right hand side of
Eq.~(\ref{E2}) goes to infinity as the cut-off $\Lambda \to \infty $.
For the summation on the right hand side of Eq.~(\ref{energyb}), 
if the value of $|d_{\beta '}(p-k,\Lambda ) |^2$ decreases fast enough
with increasing $k_0$, then, it would be possible for it to be convergent
in the limit of $\Lambda \to \infty $.
That is to say, for a model suffering ultraviolet divergence
in perturbation theory, when it is treated nonperturbatively,
it is possible for some of, even all of, its eigenstates to have
finite eigenenergies. 

    At last, we would like to mention that the theory of relativity
may give additional restrictions to possible physical eigenstates
with finite eigenenergies in the limit of $\Lambda \to \infty $.
For example, the requirement that $E_{\beta }(p, \infty )$
is finite does not guarantee that it satisfies
the relation
\be E_{\beta }^2(p,\infty ) = E_{\beta }^2(0,\infty ) +p^2.
\label{EP} \ee
Furthermore, the theory of relativity requires
that Lorentzian transformation can transform an eigenstate
$|\phi_{\beta}(p,\infty ) \rangle $ to an eigenstate
$|\phi_{\beta }(p',\infty ) \rangle $. But, it is not clear if all
the eigenstates given in Eq.~(\ref{psib}) 
satisfy this requirement.

\subsection{Evolution of states in infinite Hilbert space }
\label{sect4.3}

    Finally, let us give a brief discussion for evolution of states
in the infinite Hilbert space $L_{\infty }(p,\lambda )$.
In the infinite Hilbert space, the subspace spanned by 
the eigenstates $|\phi_{\beta }(p,\Lambda )\rangle $ of the Hamiltonian
$H(\Lambda )$, which will be called {\it eigen-subspace },
play a special role. 
States in the eigen-subspace can be expanded in the
eigenstates $|\phi_{\beta }(p,\Lambda )\rangle $, therefore,
they evolve in the same way as those in a finite
Hilbert space, say, for an initial
state $|\Phi (t=0)\rangle = |\Phi _0\rangle $,
\be |\Phi(t)\rangle = \sum_{p,\beta }
\langle \phi_{\beta }(p,\Lambda )|\Phi_0\rangle
e^{-iE_{\beta }(p,\Lambda )t} |\phi_{\beta }(p,\Lambda )\rangle .
\label{psit} \ee

    In the general case, evolution of states is more complicated than
that in Eq.~(\ref{psit}). For example, consider
a state $|\Psi _0 \rangle $
lie in a finite subspace $L_{n}(p,\lambda )$ of the infinite
Hilbert space $L_{\infty }(p,\lambda )$.
It can be divided into two parts: one in the eigen-subspace,
denoted by $|\Psi_{0}^{es} \rangle$, the other out side of the
eigen-subspace, denoted by $|\Psi_0^{nes} \rangle $.
Then, the $|\Psi_{0}^{es} \rangle$ part of the state will
evolve as in Eq.~(\ref{psit}).
But, the evolution of $|\Psi_{0}^{nes} \rangle$ is not so clear.
As time increases, it is possible for it to spread to
subspaces of the Hilbert space spanned by states in the sets
$\{ H_I^m(\Lambda )|f_p\rangle \} $ for whatever large $m$.
In this case, when we are concerned with properties of the state
in a finite subspace $L_n(p,\Lambda )$ only,
the probability of the state $|\Psi^{nes}(t) \rangle$ 
in the subspace $L_n(p,\Lambda )$
will become smaller and smaller as $t$ increases,
which remind one of properties of dissipative systems.
At the present stage, it is not clear if this feature of
evolution of states is common to 
all the states out side of the eigen-subspace.

\section{Conclusions and discussions}
\label{sect5}

    In this paper, we study a simple quantized model, which has an interaction
structure similar to (but simpler than ) that of QED
and gives ultravioletly divergent results 
in the framework of perturbation theory.
We show that when the eigen-problem of the Hamiltonian
of the model is treated nonperturbatively, 
it is in fact possible for
eigenenergies of the Hamiltonian to be finite.
The eigenstates of the Hamiltonian are found to 
span part of the whole infinite Hilbert space only.
Evolution of states in the infinite Hilbert space 
shows features quite different from that in finite Hilbert spaces.

    We expect that the method introduced in this paper is also useful in
studying more realistic models, such as the standard model and
models including the gravity.
In the application of this method to 
realistic models, masses of particles are to be explained
as energies of ground states of interacting fields, therefore,
there would be no need to resort to Higgs mechanism to get
masses for particles. 
For the standard model, without introducing
masses to free fields,
the Hamiltonian of the model can not have any non-zero
finite eigenenergy due to the lack of the dimension of mass.
But, making use of the method discussed in this paper,
it would be possible for one to get information on ratios of the
eigenenergies and properties of the eigenstates of the Hamiltonian
of the standard model.
In order to obtain non-zero finite eigenenergies
without introducing masses to free fields, one must include
the gravity. One of the most serious problems one meets
when including the gravity in quantum field theories is that the
extended theories are generally non-renormalizable.
However, as discussed in this paper, it is not impossible
for non-renormalizable models to have energy eigenstates
with finite energies when treated rigorously. Therefore, 
besides the string theory (see, e.g., \cite{Poly,RMP}),
the method introduced in this
paper may supply another possible way of overcoming the difficulty
of ultraviolet divergence.

%\end{multicols}

\end{document}